\newcommand{\SU}{SU}

\documentclass[%
 reprint,
nofootinbib,
 amsmath,amssymb,
 aps,
prb,
]{revtex4-2}

\usepackage[caption=false]{subfig}

\usepackage[dvipsnames]{xcolor}
\usepackage{graphicx}
\usepackage{dcolumn}
\usepackage{bm}

\usepackage{slashed}
\usepackage{mathtools}
\usepackage{cleveref}

\begin{document}


\title{Heavy Dark Baryons at Large \(N\): Self-Interactions across All Scales}

\author{Giovani Dalla Valle Garcia}%
 \email{giovani.dallavallegarcia@unimelb.edu.au}
\affiliation{ARC Centre of Excellence for Dark Matter Particle Physics,
School of Physics, The University of Melbourne, Victoria 3010, Australia}
\author{Juan Herrero-Garc\'{\i}a}%
\affiliation{Departament de Física Teòrica, Universitat de València, 46100 Burjassot, Spain}
\affiliation{Instituto de Física Corpuscular (CSIC-Universitat de València),
Parc Científic UV, C/Catedrático José Beltrán, 2, E-46980 Paterna, Spain\vspace{0.075cm}}

\date{\today}
\begin{abstract}
We show that heavy dark baryons in a confining $\SU(N)$ sector with a
moderately large number of colors, $N\sim10$--$100$, can generate self-interactions spanning
the full range of astrophysical velocities. Their chromo-electric
polarizability induces van der Waals interactions that yield
$\sigma/m_{\rm DM}\sim30$--$100~\mathrm{cm^2/g}$ at dwarf-galaxy
velocities, decrease to the values required by galaxy clusters, and
rise to $10^2$--$10^3~\mathrm{cm^2/g}$ at velocities of a few
$\mathrm{km/s}$, as suggested by the second perturber of JVAS
B1938+666.  Remarkably, for a given ratio $m_Q/\Lambda_d$, these
observations can be used to infer the value of $N$.  The resulting self-interaction phenomenology favors GeV-scale dark baryons, independently of the relic-abundance
requirement. At these masses, a dark baryon asymmetry comparable to the
visible one can account for the observed dark-matter abundance.
\end{abstract}
 
\maketitle


\section{\label{sec:intro}Introduction}Self-interacting dark matter (SIDM) can reconcile small-scale structure problems with an otherwise successful cold-dark-matter cosmology~\cite{Spergel:1999mh,Tulin:2017ara,Bullock:2017xww}. In particular, recent works highlight gravothermal evolution in the strong-SIDM regime as a possible origin of the observed diversity of galactic rotation curves~\cite{Zavala:2019sjk,Correa:2020qam,Correa:2022dey,Roberts:2024uyw}. The required interaction is sharply scale dependent: dwarf spheroidals favor $\sigma/m_{\rm DM}\sim30$--$100~{\rm cm}^2/{\rm g}$ near $30~{\rm km/s}$~\cite{Correa:2020qam}, whereas clusters require $\sigma/m_{\rm DM}\lesssim1~{\rm cm}^2/{\rm g}$ near $1500~{\rm km/s}$~\cite{Correa:2020qam,Balan:2024cmq}.  Strong lensing may pose an even more demanding target. The dense second perturber of JVAS B1938+666 has been interpreted as a gravothermally collapsed subhalo, requiring $\sigma/m_{\rm DM}\sim10^2$--$10^3~{\rm cm}^2/{\rm g}$ at a few $\mathrm{km/s}$~\cite{Powell:2025rmj,Vegetti:2026mmx,Zhang:2026gur}. The challenge is therefore to obtain sizable self-interactions with a strength that varies by orders of magnitude across astrophysical velocities.

\begin{figure}[t]
\centering
\includegraphics[width=1\linewidth]{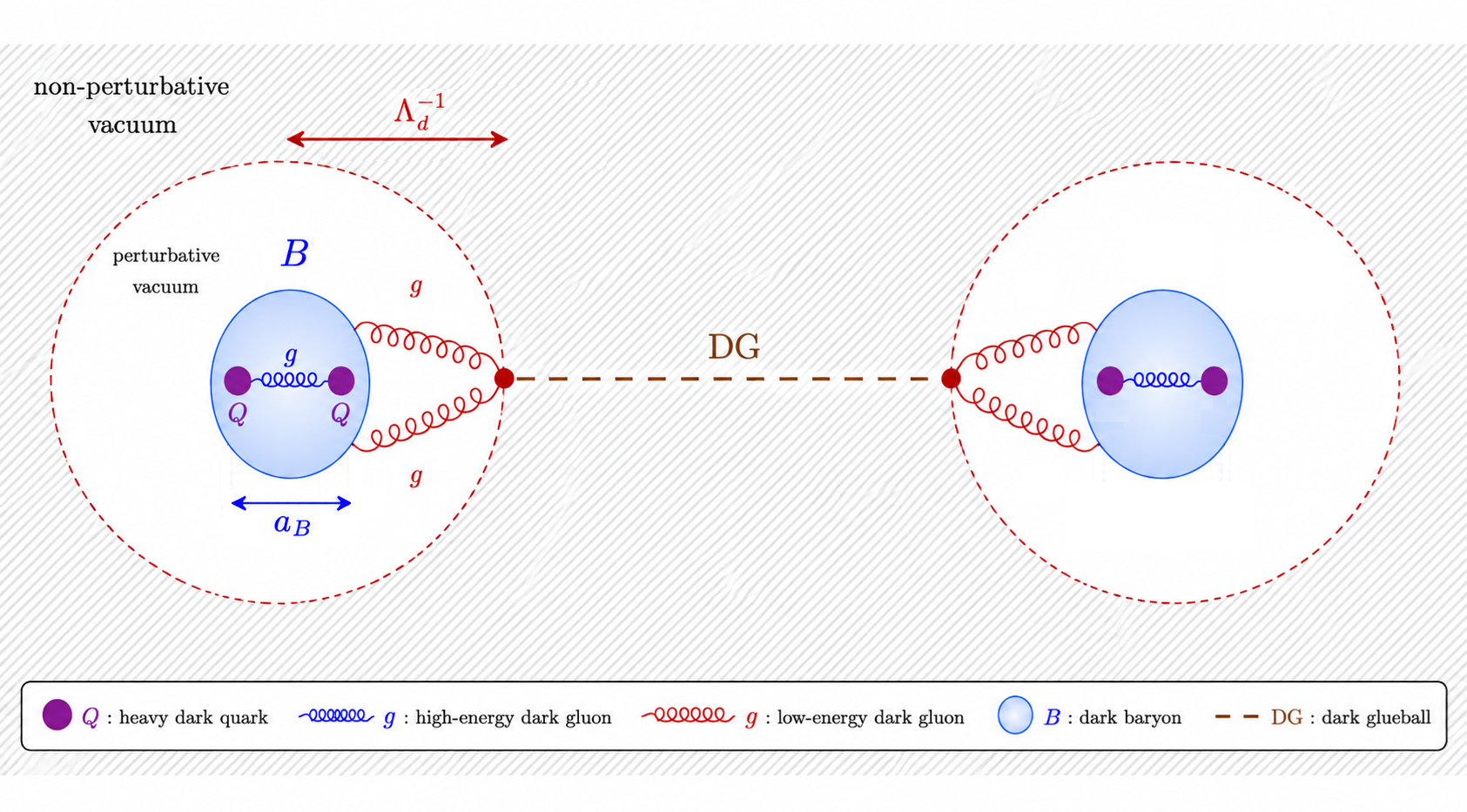}
\caption{Schematic illustration of the interaction between two dark baryons in an \(\SU(2)\) confining gauge theory. Each baryon consists of two heavy quarks \(Q\), and the compact color-singlet states interact through induced chromo-electric dipoles, while their short-distance overlap generates a repulsive core.}
\label{fig:SIDM}
\end{figure}

We show that such a velocity dependence can arise from the structure of a
minimal heavy confining sector: an $\SU(N)$ gauge theory with a single
vector-like quark of mass larger than the confinement scale,
$m_Q\gg\Lambda_d$. Its lightest baryon is a stable $N$-quark
color singlet and a compact Coulombic state. As a color singlet, it does not experience a
one-gluon-exchange force. Instead, its chromo-electric polarizability induces
a non-Abelian van der Waals interaction, analogous to that between neutral
atoms and to heavy-quarkonium interactions in QCD
~\cite{Peskin:1979va,Bhanot:1979vb,Fujii:1999xn}. This is illustrated
schematically in \cref{fig:SIDM} for the case of $N=2$ for simplicity.
The force crosses from a London dispersion to a retarded Casimir--Polder
form and finally to a glueball Yukawa tail, while fermion exchange supplies
a repulsive core. Its velocity dependence is thus a consequence of
compositeness and confinement. This approach differs significantly from previous estimates~\cite{Boddy:2014qxa,Boddy:2014yra}, which assume an $\mathcal{O}(1)$ coupling for the glueball-mediated Yukawa potential, whereas here its strength is fixed by the chromo-electric polarizability of the heavy baryon.

For moderately large $N$, the heavy-baryon interaction approaches an approximately $N$-independent behavior. We use $\SU(20)$ as a representative finite-$N$ realization. Although the precise location of the resonant-like enhancement depends on $N$, its qualitative velocity dependence persists for $N\sim7$--$80$, with the corresponding features occurring at $m_Q/\Lambda_d\sim10$--$100$; see Appendix~\ref{app:dependence_on_N}.

Unlike glueball dark matter, whose interactions are controlled mainly by
$\Lambda_d$ and are consequently only mildly velocity dependent
~\cite{Boddy:2014yra,Yamanaka:2019gak,Yamanaka:2019yek}, heavy baryons introduce a second
dynamical scale through $m_Q$. Light-quark dark hadrons can also generate
strong velocity dependence, but typically require several quark flavors,
CP-violating dynamics, or very narrow resonances
~\cite{Cline:2022leq,Garcia-Cely:2024ivo,Figueroa:2026bmi}.
Here it follows from $m_Q\gg\Lambda_d$ with only one quark flavor and no
extra elementary mediator~\cite{Chung:2025wle}.

Solving the scattering problem, we find that the large-$N$ regime contains
regions of parameter space that simultaneously satisfy the dwarf-scale
self-interaction requirements, obey cluster limits, and reach the strength
associated with JVAS B1938+666. These combined self-interaction requirements select
$m_B\sim 10m_p$ and $\Lambda_d\sim\Lambda_{\rm QCD}$. In this same region, annihilations
efficiently remove the symmetric baryon population, so that a dark
asymmetry comparable to the visible-baryon asymmetry yields the
observed dark matter abundance.

In this Letter, we present the essential ingredients and main results of
this scenario; detailed derivations are given in the companion work
~\cite{DallaValleGarcia:2026companion}. The relevant properties of our 
confining dark sector and its heavy Coulombic dark baryons are introduced in Sec.~\ref{sec:heavy_dark_baryons}. Sec.~\ref{sec:interaction_potential}
discusses the different regimes of the baryon--baryon interaction potential and the short-distance regularization adopted in our analysis.
Sec.~\ref{sec:self_interactions} introduces the relevant self-interaction
observables and summarizes the numerical procedure used to compute the
scattering cross sections. Sec.~\ref{sec:results} presents the resulting
velocity-dependent self-interactions and their implications across the
relevant astrophysical scales. Sec.~\ref{sec:asymmetric_DM} discusses the
cosmological implications and the asymmetric origin of the dark-baryon
abundance. We conclude in Sec.~\ref{sec:conclusion}.

\section{\label{sec:heavy_dark_baryons}Heavy Dark Baryons}

We consider a confining $\SU(N)$ gauge theory with gauge coupling $g_d$
and a single vector-like fermion $Q$ in the fundamental representation.
The dark fermion is neutral under the Standard Model and has mass $m_Q$. We define
\[
    \alpha_d \equiv \frac{g_d^2}{4\pi}
\]
and denote the confinement scale by $\Lambda_d$. We focus on the
heavy-quark regime $m_Q\gg\Lambda_d$, in which the velocity dependence
of the baryonic self-interactions is controlled only by the ratio
of scales 
\[
    \xi\equiv\frac{m_Q}{\Lambda_d}.
\]

At temperatures below the confinement scale $\Lambda_d$, the spectrum
contains glueballs as well as heavy mesons and baryons. The heavy mesons
are not protected by a conserved quantum number and can promptly decay
into glueballs, whereas the lightest baryon is stable due to the
accidental dark baryon number. Glueballs are likewise not protected by
an exact conserved quantum number, and their masses satisfy
$m_{\rm DG}\sim\Lambda_d\ll m_B\simeq Nm_Q$, such that suitable portals to the SM
or a colder dark sector can deplete their abundances. We therefore
identify the lightest baryon, composed of $N$ heavy quarks, as the
dark-matter candidate.

In the heavy-quark limit $m_Q\gg\Lambda_d$, the baryon is a compact
Coulombic bound state~\cite{Mitridate:2017oky}. This means that the
linear confining potential can be neglected, while pairs of quarks
within the color-singlet baryon interact through the Coulomb potential (taking $N\gg1$)
\begin{equation}
     V_s(r_{ij})=
    -\frac{1}{2}\frac{\alpha_d}{r_{ij}}\,,
\end{equation}
where $r_{ij}$ is the distance between quarks $i$ and $j$. For
large $N$, the baryon contains a large number of quarks and can
be accurately described in the mean-field approximation. The resulting
characteristic Bohr radius and binding energy per quark, which we denote
by $a_B$ and $\epsilon_B$, respectively, are then given by
~\cite{Cohen:2011cw}
\begin{align}
    a_B^{-1}&\simeq 0.6\tilde\alpha_d m_Q,
    &
    \epsilon_B&\simeq0.05\tilde\alpha_d^2m_Q\,,
\end{align}
where $\tilde\alpha_d\equiv N\alpha_d$  remains finite and independent of $N$ in the
large-$N$ limit.

The gauge coupling entering these quantities is evaluated at the
characteristic momentum scale of the bound state, $p\simeq a_B^{-1}$.
Thus, for a given $\xi$, $\tilde\alpha_d(p)$ is determined
self-consistently from the running of $\alpha_d$, \begin{equation}
\label{eq:alpha_per_energy}
\tilde\alpha_d(p)
\simeq
\frac{6\pi}
{11\ln\!\left(0.6\tilde\alpha_d(p) \xi\right)}\,.
\end{equation} 

The Coulombic description requires the baryon size to remain
sufficiently smaller than the confinement length and the constituent
velocities to remain well below the speed of light. Estimating
confinement effects by including a linear contribution to the potential
$\Lambda_d^2 r$~\cite{Mitridate:2017oky} shows that they remain
subleading over the range of $\xi$ relevant for our analysis.
Relativistic corrections, including chromo-magnetic contributions, are
also suppressed by powers of the constituent velocity
$v\sim\tilde\alpha_d/2$. We therefore  restrict our
analysis to
\begin{equation}
    \xi\geq10\,,
\end{equation}
where both effects remain under control.

\section{\label{sec:interaction_potential}Interaction Potential}Since the dark baryons are compact color singlets, their leading interaction
at distances larger than their size is not given by one-gluon exchange.
Instead, a chromo-electric fluctuation can excite one baryon into a color
adjoint dipole state, whose chromo-electric field polarizes the second baryon.
The correlated two-gluon exchange then returns both baryons to their
color-singlet ground states, generating an attractive induced
dipole--dipole interaction. 

The detailed derivation of the interaction, including the spectral
representation of the two-gluon-induced chromo-electric fields, the
chromo-electric polarizability using an approximate treatment of intermediate adjoint states, and the matching of the perturbative and
non-perturbative spectra, is presented in the companion work~\cite{DallaValleGarcia:2026companion}.
For the large-$N$ regime considered here, the resulting long-distance
potential has three distinct regimes and is approximately given by 
\begin{widetext}
\begin{equation} \label{eq:Vr_definition}  
V(r)\approx-\frac{\tilde d_2^2 a_B^6}{16\pi^2}
\begin{cases}
\displaystyle \frac{21}{32}\epsilon_B r^{-6}\,,  & \qquad
a_B\ll r \lesssim \epsilon_B^{-1}\,,\\[3mm]
\displaystyle \frac{23}{4\pi}r^{-7}
 \,, &\qquad
\epsilon_B^{-1} \lesssim r \lesssim\Lambda_d^{-1}\,,\\[2mm]
\displaystyle C_Y\,\frac{e^{-m_{\rm DG}r}}{r}\,, &\qquad
r\gtrsim \Lambda_d^{-1}\,,
\end{cases}
 \end{equation} 
\end{widetext}
where the reduced chromo-electric polarizability coefficient is
$\tilde d_2\simeq160$,  and
\begin{equation}
C_Y =
\frac{1}{\tilde\alpha_d^2}
\frac{4\pi^3}{3025}
m_{\rm DG}^6\,,
\end{equation}
using the glueball spectral density computed on the lattice for $SU(3)$~\cite{Chen:2005mg,Meyer:2008tr} and its large-\(N\) scaling.\footnote{For sufficiently small $m_Q/\Lambda_d $, one can have
$ \epsilon_B \lesssim\Lambda_d$, in which case the Casimir--Polder regime is
absent. In these cases, non-perturbative effects could also penetrate below $\Lambda_d^{-1}$ and $\epsilon_B^{-1}$;
we therefore add the Yukawa  contribution on top of the perturbative potentials down to 
$r\geq r_{\rm NP}=0.4/\epsilon_B$. The associated uncertainty is discussed
in the companion work~\cite{DallaValleGarcia:2026companion}.} Note that the glueball mass $m_{\rm DG}\simeq 6.4\Lambda_d$~\cite{Teper:1998kw,Gockeler:2005rv} up to small $\mathcal{O}(1/N^2)$ corrections~\cite{Forestell:2016qhc}. The full potential for generic $N$ used in the analysis is given in Ref.~\cite{DallaValleGarcia:2026companion}.

The first regime is the non-retarded London interaction, arising from
correlated dipole fluctuations. At distances larger than the inverse
binding energy, retardation becomes important and gives the
Casimir--Polder $r^{-7}$ potential, precisely as for ordinary hydrogen
~\cite{Casimir:1947kzi,Dzyaloshinskii1956}. Finally, at distances of
order the confinement length and beyond, the interaction is dominated
by exchange of the lightest dark glueball and becomes fully
Yukawa-like. The overall strength of the dipole-induced interaction is
fixed by the chromo-electric polarizability of the Coulombic baryon,
directly related to $\tilde d_2$, whose spectral sum is derived in the
companion work~\cite{DallaValleGarcia:2026companion}.

\subsection{\label{sec:regularizing_potential}Short-Distance Regularization}

The long-distance potential cannot be extrapolated to separations
$r\to0$, where the multipole expansion breaks down and the constituent
wave functions begin to overlap. We therefore suppress the attractive
potential below a matching radius $r_{\rm in}$~\cite{Cline:2013pca},
\begin{equation}\label{eq:regularized_potential}
V_{\rm reg}(r)
= e^{-\left(\frac{r_{\rm in}}{r}-1\right)^2}V(r)
\qquad
\text{for } r\leq r_{\rm in},
\end{equation}
and include the effective Pauli repulsive potential
\begin{equation}
    V_{\rm rep}(r)
     \approx 11.6\,N\,\epsilon_B
     \left(\dfrac{r}{a_B}\right)^7
     e^{-6.4r/a_B}\,,
\label{eq:Vrep_letter}
\end{equation}
computed in the large-$N$ regime~\cite{Adhikari:2013dfa}. The total
potential is then
\begin{equation}\label{eq:total_potentail}
    V_{\rm tot}=V_{\rm reg}+V_{\rm rep}\, .
\end{equation}
We adopt the relatively conservative benchmark $r_{\rm in}=7a_B$ and
denote the corresponding potential by $V_{\rm tot7}$. Varying
$r_{\rm in}$ from $a_B$ to $\sim10a_B$ shifts the resonant values of
$\xi$ but generally preserves the enhancement and high-velocity
suppression central to our result. The microscopic motivation for this
regularization and the associated uncertainty are discussed in the
companion work~\cite{DallaValleGarcia:2026companion}.

An important feature of the large-$N$ potential is its different
scaling with $N$ in the attractive and repulsive contributions. The
attractive potential becomes approximately $N$ independent, up to
$\mathcal{O}(1/N^2)$ corrections associated with the large-$N$
't Hooft scaling of $\tilde\alpha_d$, whereas the Pauli repulsion grows
linearly with $N$ as the number of overlapping constituents increases.
Consequently, when the repulsive contribution remains subdominant, the
scattering dynamics is only mildly sensitive to $N$, apart from the
explicit dependence of the baryon mass $m_B\simeq Nm_Q$. This is the
regime relevant for the moderately large values of $N$ considered here.
As $N$ increases further, however, the Pauli repulsion eventually
dominates the interaction, driving the cross section toward a
geometric-like behavior and suppressing the strong resonant features
needed for gravothermal core-collapse signatures. The resulting
dependence on $N$ is explored in Appendix~\ref{app:dependence_on_N}.

\section{\label{sec:self_interactions}Self-Interactions}

As shown below, the dark matter relic is asymmetric. The relevant
self-scattering therefore involves identical dark baryons, for which
momentum redistribution is characterized by the viscosity cross section
\begin{equation}
    \sigma_V =
    \int d\Omega\,\sin^2\theta\,
    \frac{d\sigma}{d\Omega}\,,
\label{eq:sigmaV}
\end{equation}
which suppresses physically ineffective forward and backward scattering
~\cite{Tulin:2013teo}.

In the one-flavor $\SU(N)$ theory considered here, the ground-state
baryon has spin $S=N/2$. Averaging over the initial spin states gives
different weights for even and odd partial waves, depending on whether
the baryon is a boson or a fermion. In the large-$N$ limit, however,
these weights approach $1/2$ for both parity classes, and the viscosity
cross section becomes simply
\begin{equation}
    \sigma_V
    \simeq
    \frac{4\pi}{k^2}
    \sum_{\ell}
    \frac{(\ell+1)(\ell+2)}{2\ell+3}
    \sin^2\left(\delta_{\ell+2}-\delta_\ell\right).
\label{eq:sigmaV_partial}
\end{equation}
Thus, at leading order in the large-$N$ limit, the result is independent
of whether the dark baryons are bosonic or fermionic.

The phase shifts $\delta_\ell$ are obtained by solving the radial
Schr\"odinger equation for the reduced wave function $u_\ell(r)$ in the
full regularized potential $V_{\rm tot7}$. Given the reduced mass
$\mu=m_B/2$ and relative momentum $k=\mu v$, the equation reads
~\cite{Buckley:2009in,Tulin:2013teo}
\begin{equation}\label{eq:radial_schr_normalized}
    u_\ell''(x)
    +
    \left(
        a^2
        -
        \frac{\ell(\ell+1)}{x^2}
        -
        2\,\mu\,a_B[a_B V_{\rm tot7}(x)]
    \right)
    u_\ell(x)
    =
    0\,,
\end{equation}
where we define the dimensionless variables
$x\equiv r/a_B$ and
$a\equiv ka_B=v\mu a_B$. At large $r$, where the potential becomes
negligible, the numerical solution approaches
\begin{equation}
    u_\ell(r)\xrightarrow[r\to\infty]{}
    A_\ell\sin\!\left(kr-\frac{\ell\pi}{2}+\delta_\ell\right),
\end{equation}
which defines the phase shift $\delta_\ell$.

For fixed $\xi$, the variables $a$, $\mu a_B$, and $a_BV_{\rm tot7}(x)$ are independent of the overall quark-mass scale. Thus, changing $m_Q$ leaves the velocity dependence of the cross section unchanged while modifying its overall normalization. Moreover, $\mu a_B\sim N/\tilde\alpha_d$, so the scattering retains an explicit $N$ dependence even when the reduced potential is approximately $N$ independent. In particular within the Born approximation, this gives the parametric scaling $\sigma_{\rm Born}\propto N^2$.

Astrophysical systems probe a distribution of relative velocities rather
than a single velocity. We therefore define the velocity-averaged
viscosity cross section
\begin{equation}
    \overline{\sigma}_V
    =
    \frac{\langle \sigma_V v^\gamma\rangle}
    {\langle v^\gamma\rangle}
    =
    \frac{
    \int_0^\infty dv\,
    f_{\rm rel}(v)\,
    v^\gamma\,\sigma_V(v)
    }{
    \int_0^\infty dv\,
    f_{\rm rel}(v)\,
    v^\gamma
    },
\label{eq:velocity_averaged_cross_section}
\end{equation}
where $f_{\rm rel}$ is the Maxwell--Boltzmann relative-velocity
distribution and $\gamma=1$ ($3$) corresponds to rate
(energy-transfer) weighting~\cite{Kaplinghat:2015aga,Ramos:2025lvk}.
We use representative velocities for the parameter scan and both
weightings for the benchmark point.

We evaluate the cross section in the regime where scattering remains
elastic and the internal structure of the baryons is not directly
resolved. The latter requires the shortest wavelength probed in the
collision to remain larger than the baryonic size,
\begin{equation}
    k_{\rm max}^{-1}> a_B\,,
\label{eq:kmax_condition}
\end{equation} where $k=\mu v_{\rm max}$ with $v_{\rm max} \sim 1500~{\rm km/s}$ being the largest velocity relevant for the astrophysical systems considered. Elasticity further requires the maximum centre-of-mass kinetic energy
to remain below both the baryon binding energy and the lightest
glueball mass,
\begin{equation}
    E_{\rm max}
    \simeq \frac{\mu v_{\rm max}^2}{2}
    \ll \epsilon_B,
    \qquad
    E_{\rm max}<m_{\rm DG}\,.
\label{eq:elasticity_condition}
\end{equation}
These conditions ensure that the baryons are neither resolved nor
excited or dissociated, and that dark-glueball production is
kinematically forbidden. They are satisfied throughout the parameter
space relevant for the self-interaction phenomenology presented below.
Further details on the partial-wave treatment for generic $N$,
convergence tests, and the numerical implementation are given in the
companion work~\cite{DallaValleGarcia:2026companion}.

\begin{figure}[t]
             \centering \includegraphics[width=1\linewidth]{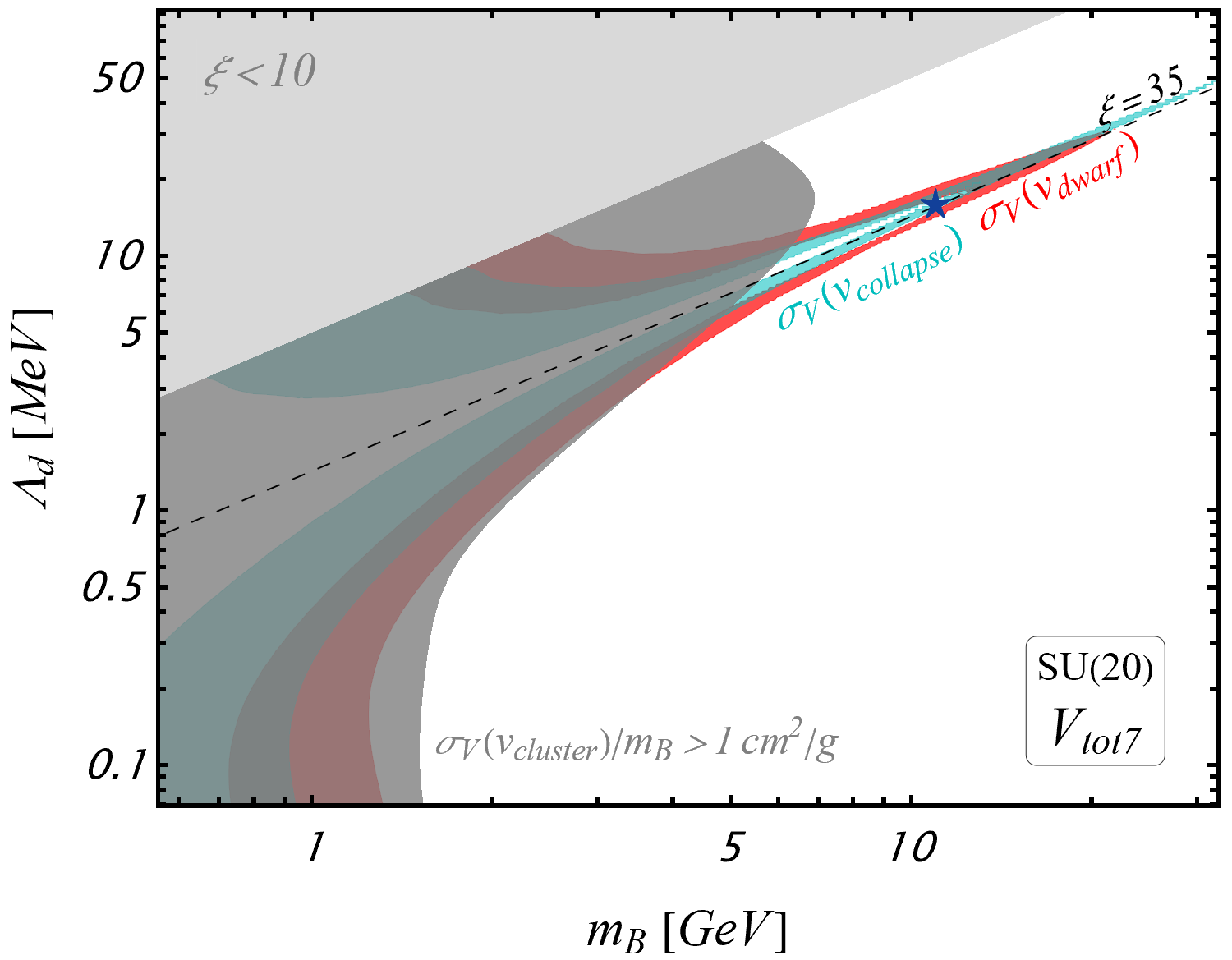}

            \caption{Dark-baryon viscosity self-interaction cross section per unit mass in
the $(m_B,\Lambda_d)$ plane, computed using $V_{\rm tot7}$ for $SU(20)$. Red
regions yield $30\;\mathrm{cm}^2/\mathrm{g}\leq\sigma_V/m_B\leq100\;\mathrm{cm}^2/\mathrm{g}$
at $v_{\rm dwarf}=30~\mathrm{km/s}$~\cite{Correa:2020qam}. Dark-gray regions violate the cluster
constraint $\sigma_V/m_B\lesssim 1\;\mathrm{cm}^2/\mathrm{g}$ at
$v_{\rm cluster}=1500~\mathrm{km/s}$~\cite{Balan:2024cmq}, while light-gray regions lie outside
the domain of validity of our theoretical treatment. Light blue regions satisfy
$75\leq\sigma_V/m_B\leq1000~\mathrm{cm}^2/\mathrm{g}$ at
$v_{\rm collapse}=5~\mathrm{km/s}$, as suggested by the second perturber of
JVAS B1938+666~\cite{Vegetti:2026mmx,Zhang:2026gur}. The dashed black line
marks the benchmark choice $\xi=35$ used to study the velocity dependence in
detail, while the $\star$-point is the best fit $(m_B,\Lambda_d)$ along the $\xi=35$ line presented in \cref{fig:final_plot}.}
    \label{fig:plane_mqXlambda}
\end{figure}

\section{Results}\label{sec:results}For fixed $N$, the velocity dependence of the self-interaction cross section is controlled entirely by the hierarchy
$\xi=m_Q/\Lambda_d$. If we also fix $\xi$, then the dimensionless combination
$m_B^3(\sigma_V/m_B)$  is uniquely determined, while the overall normalization can be recovered by rescaling the dark-quark mass. This allows us to efficiently scan the $(m_B,\Lambda_d)$ parameter space by first computing the cross section as a function of $\xi$ and subsequently rescaling the baryon mass. 

The resulting parameter space for $SU(20)$ is shown in \cref{fig:plane_mqXlambda}. A common region satisfies the self-interaction requirements across the different astrophysical scales, exhibiting strong interactions in dwarfs, suppression at cluster velocities, and substantial enhancement at lower velocities.

\begin{figure}[t]
\includegraphics[width=1\linewidth]{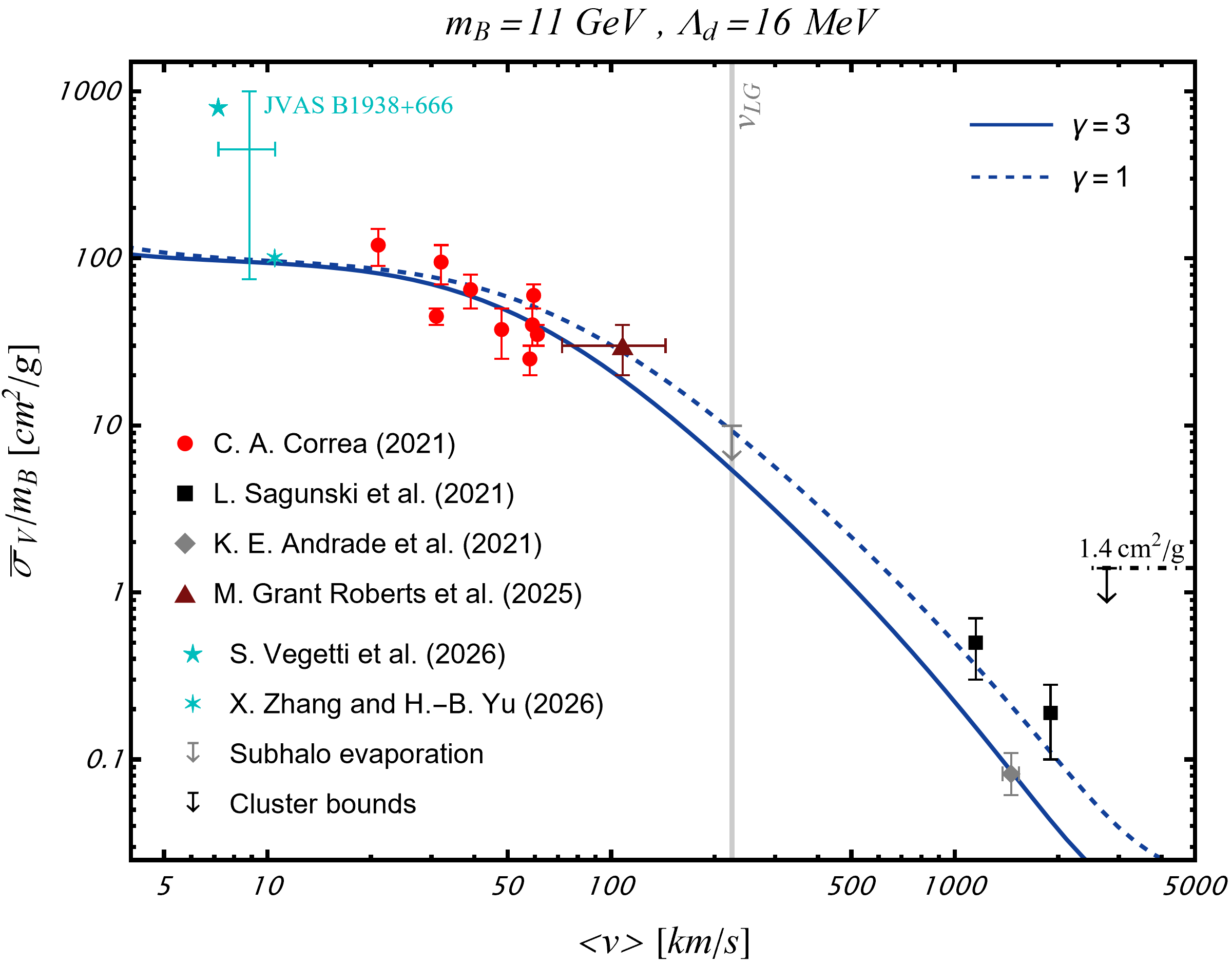}
\caption{Velocity-averaged viscosity cross section per unit mass for the $SU(20)$ benchmark point in \cref{fig:plane_mqXlambda}, computed with $V_{\rm tot7}$. Black and gray points show inferences from galaxy clusters and groups~\cite{Sagunski:2020spe,Andrade:2020lqq}; light red points show dwarf-spheroidal determinations~\cite{Correa:2020qam}, while dark red ones show low surface brightness galaxy results~\cite{Roberts:2024uyw}; and light blue intervals show the approximate range inferred for the second perturber of JVAS B1938+666~\cite{Vegetti:2026mmx,Zhang:2026gur}. Dashed and solid curves denote the rate-weighted ($\gamma=1$) and energy-transfer-weighted ($\gamma=3$) averaged cross sections for the best fit. We also show the Local Group evaporation~\cite{Correa:2020qam} and the Bullet Cluster bounds~\cite{Balan:2024cmq}. } 
\label{fig:final_plot}
\end{figure}

To illustrate the resulting velocity dependence, we select the benchmark $\xi=35$ which crosses the region where the different requirements overlap and then fix its
normalization by determining $m_B$ through a fit that---to ensure robustness---relies exclusively on inferences derived from dwarf spheroidal galaxies~\cite{Correa:2020qam}.  The result of this fit yields $m_B=11$ GeV, corresponding to $\Lambda_d=16$ MeV. Its prediction over the
full range of astrophysical velocities is shown in \cref{fig:final_plot}.
The two curves correspond to the rate-weighted and energy-transfer-weighted averages defined in \cref{eq:velocity_averaged_cross_section}.
Despite the different weighting prescriptions, the qualitative behavior is
unchanged: in both cases the cross sections exhibit a rapid decrease from dwarf to cluster
velocities, satisfying the cluster bound, while at lower
velocities they increase sufficiently to reach the range inferred for the
second perturber of JVAS B1938+666. At Local Group velocities,
$v_{\rm LG}\simeq225~\mathrm{km/s}$, both prescriptions have already
decreased sufficiently to satisfy the
$\sigma/m<10~\mathrm{cm}^2/\mathrm{g}$ bound due to the absence of significant
subhalo evaporation
~\cite{Correa:2020qam,Vogelsberger:2012ku,Rocha:2012jg,Zavala:2012us}.
Fixing $m_B$ solely from dwarf-spheroidal data also yields cluster-scale cross sections in the range preferred by the cluster determinations, rather than merely satisfying the corresponding upper bound. 
 
Remarkably, the region satisfying these requirements corresponds to dark-baryon masses and confinement scales not far from the familiar QCD scales. This coincidence provides a natural setting for an asymmetric dark matter origin, which we discuss in the following.

 \section{Asymmetric origin}\label{sec:asymmetric_DM}

The parameter region favored by the self-interaction phenomenology has
$m_B\sim\mathcal{O}(10~\mathrm{GeV})$ and
$\Lambda_d\sim\mathcal{O}(10~\mathrm{MeV})$. These scales also have important
implications for the cosmological origin of the dark matter. For these
relatively light dark baryon masses and moderate hierarchy ratios
$\xi\sim30$, the symmetric population of heavy dark quarks efficiently
annihilates through $Q\bar Q\rightarrow gg$ down to the confinement
temperature $T_d\sim\Lambda_d$. Moreover, after confinement,
baryon--antibaryon rearrangement into mesons leads to further efficient
reannihilation~\cite{Mitridate:2017oky}, removing essentially the entire
symmetric component~\cite{DallaValleGarcia:2026companion}.
Bound-state formation~\cite{Binder:2023ckj,Flores:2026yay} and effects
associated with the confinement phase transition~\cite{Gouttenoire:2023bqh}
can further reduce the symmetric abundance. The surviving symmetric
component is therefore negligible in the parameter region relevant for the
self-interactions.

The remaining dark baryons can naturally be generated by a primordial dark
baryon asymmetry. Since the lightest dark baryon is stabilized by the
accidental dark baryon number of the minimal theory, an asymmetry
$\eta_D$ can survive the efficient annihilation of the symmetric component.
For the ${O}(10~\mathrm{GeV})$-scale dark baryon masses selected by the self-interaction
phenomenology, an asymmetry of the same order as the visible baryon
asymmetry naturally gives the observed dark matter abundance
\cite{Murgui:2021eqi,Chung:2024dgu}. The microscopic origin of $\eta_D$ is
model dependent and is not specified here.

The dark glueball abundance provides an additional cosmological constraint.
For sufficiently low $\Lambda_d$, dark gluons hadronize into glueballs whose
relic abundance can be excessive if $T_d\sim T_{\rm SM}$. This can be avoided
if the dark sector is sufficiently cold,
$T_d\lesssim10^{-3}\,T_{\rm SM}$
\cite{Carenza:2022pjd,Carenza:2023eua,DallaValleGarcia:2026companion},
or if the glueballs decay through a suitable portal. Since they generally possess no stabilizing symmetry, such decays are not forbidden by a  conserved dark quantum number.\footnote{Note, however, that for $N \geq 3$, it is possible to stabilize the lightest $C$-odd glueball by forbidding $C$-breaking portals~\cite{McKeen:2024trt}. } 

The absence of light quarks also makes the confinement transition first order, potentially sourcing gravitational waves \cite{YMtransition}. Although our benchmark suggests \(T_c=\mathcal{O}(\Lambda_d)\sim10\!-\!20\,\mathrm{MeV}\), a Pulsar Timing Array (PTA) interpretation requires a sufficiently strong and slow transition \cite{NANOGravNP,NANOGravNPErratum}. Recent lattice-informed Yang–Mills calculations favour only modest supercooling and \(\beta/H_*\sim10^4\) near \(N=10\!-\!20\), yielding weak signals \cite{YMsupercooling,YMGW}. Moreover, the cold stable-glueball history, \(T_d/T_{\rm SM}\sim10^{-3}\), places the transition at \(T_{\rm SM}\sim10\!-\!20\,\mathrm{GeV}\) and suppresses the available dark energy fraction to order \(10^{-11}\). A warmer sector requires efficient glueball depletion and a consistent reheating history. Therefore, establishing any connection with the observed PTA background requires a dedicated cosmological analysis.

\section{Conclusion}\label{sec:conclusion}

We have shown that heavy dark baryons in a confining $\SU(N)$ gauge
theory can generate self-interactions with a strong velocity dependence.
In the moderately large-$N$ regime, the chromo-electric polarizability of
the compact Coulombic baryons produces an attractive interaction whose
interplay with the short-distance Pauli repulsion leads to pronounced resonant-like
enhancements of the self-scattering cross section. We use $\SU(20)$ as a
representative finite-$N$ realization; the qualitative velocity dependence
persists over a broad range $N\sim7$--$80$, with the enhancement occurring
for $m_Q/\Lambda_d\sim10$--$100$, although its precise location depends
on $N$.

The dependence on \(N\) also provides an additional diagnostic beyond the
resonant-like regime. For \(\xi\gg100\), the ratio of the cross sections at
collapse and cluster velocities approaches an \(N\)-dependent plateau.
Thus, if a large hierarchy \(\xi\gg100\) is established independently,
the asymptotic velocity enhancement could in principle provide information
on the number of colors. For the observational targets considered here,
this regime points to \(N\sim75\)--\(150\).

Returning to the resonant-like regime and focusing on our \(\SU(20)\) case
study, a common region of parameter space simultaneously provides
self-interactions in the range favored by dwarf spheroidal and
low-surface-brightness galaxies, satisfies galaxy-cluster constraints, and
reaches \(\mathcal{O}(100\text{--}1000)~\mathrm{cm}^2/\mathrm{g}\) at
velocities of a few \(\mathrm{km/s}\), as suggested by the second perturber
of JVAS B1938+666. The corresponding self-interaction phenomenology selects
dark baryons with masses in the tens-of-GeV range and a confinement scale
\(\Lambda_d\sim\mathcal{O}(10\text{--}100~\mathrm{MeV})\).

The same region favors an asymmetric origin for the dark matter. The
symmetric heavy-quark population is efficiently depleted, leaving the
dark-matter abundance to be set by a primordial dark baryon asymmetry.
For the GeV-scale dark baryons selected by the self-interaction
phenomenology, an asymmetry comparable to the visible baryon asymmetry
naturally yields the observed abundance. A complete cosmological
realization must additionally ensure that the dark glueball abundance is
sufficiently suppressed, for example through a colder dark sector or
suitable glueball decays. In the latter scenario, it would be interesting to study whether an observable nanohertz gravitational-wave signal can also be generated.

Our results show that strong velocity-dependent self-interactions, composite dark matter, and an asymmetric relic abundance can arise simultaneously in a minimal confining framework. The robustness of the velocity-dependent enhancement across the moderately large-$N$ regime motivates further searches for low-velocity collapsed substructures. At the same time, its sensitivity to the short-distance baryon--baryon interaction calls for more detailed calculations of the short-distance dynamics and nonperturbative effects.

\begin{acknowledgments}We are deeply indebted to Giacomo Landini for participating in the early stages of this work and for many useful discussions. We thank Bryan Zaldivar for kindly providing access to some of his numerical codes. We are also grateful to  Yi Chung, Avirup Ghosh, Pablo Figueroa, Camilo Garcia-Cely, Benjamin Grinstein, Manoj Kaplinghat, Juan Miguel Nieves, Antonio Pich, Anthony Thomas, and Hai-Bo Yu for useful discussions.  We acknowledge the
use of ChatGPT and Claude in cross-checking the
calculations and the writing. This work is partially supported by the \emph{Australian Research Council} through the ARC Centre of Excellence for Dark Matter Particle Physics (CE200100008), the Spanish \emph{Agencia Estatal de Investigación} MICINN/AEI (10.13039/501100011033) grant PID2023-148162NB-C21, the \emph{Generalitat Valenciana} through the GenT Excellence
Program (CIESGT2024-007), and the \emph{Severo Ochoa} project MCIU/AEI CEX2023-001292-S. This work has received funding from the European Union’s Horizon Europe research and
innovation programme under the Marie Skłodowska-Curie Staff Exchange grant agreement
No 101086085 – ASYMMETRY.
\end{acknowledgments}
\appendix

\bibliography{main}

\section{Evolution of the self-interaction with the number of colors}
\label{app:dependence_on_N}

Here we investigate how self-interactions evolve with the number of colors $N$
beyond the representative value considered in the main text. We focus on
$N=8$--$150$ and use the full potential $V_{\rm tot}$ with $x_{\rm in}=7$ as well as the Pauli repulsive potential $V_{\rm rep}$.
The dependence on $N$ is illustrated using two complementary diagnostics.
First, we compare the viscosity cross section obtained from the full
potential with that obtained from the repulsive core alone. This isolates
the importance of the attractive interaction relative to the increasingly
strong Pauli repulsion at large $N$. Second, we examine the ratio of the
cross sections at collapse and cluster velocities, which directly probes
the velocity dependence relevant for the phenomenology discussed in the
main text.

\begin{figure}[t]
    \centering
    \includegraphics[width=1\linewidth]{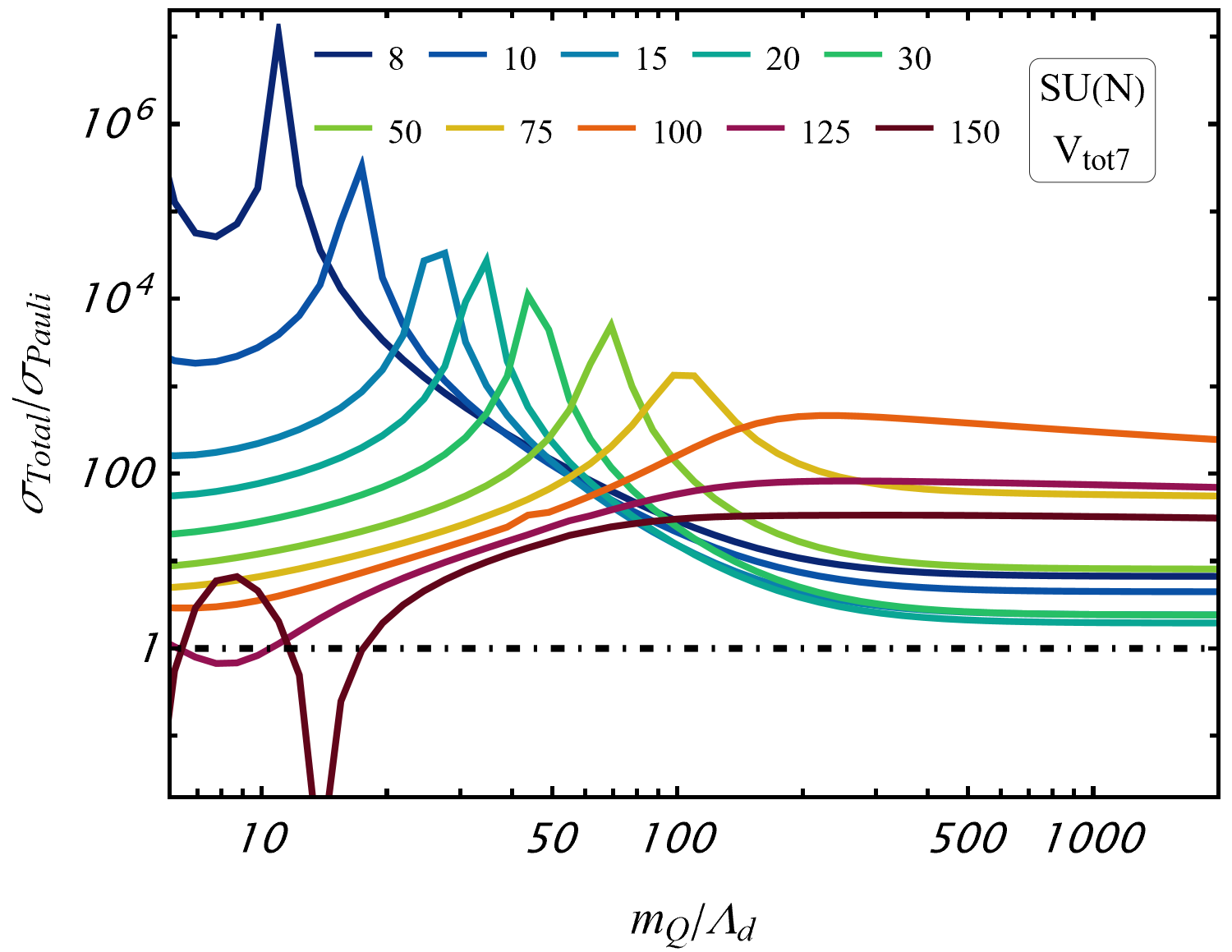}
    \caption{Ratio of the viscosity self-interaction cross section computed
    with the full potential $V_{\rm tot7}$  to that obtained using the
    repulsive potential $V_{\rm rep}$   alone, as a function of
    $\xi=m_Q/\Lambda_d$ at $v_{\rm dwarf}=30~{\rm km/s}$ for increasing values of $N$ from 8 to 150.  The pronounced enhancements for $N\lesssim75$
    correspond to the resonant-like behavior discussed in the text, while
    the broad enhancement appearing around $N\sim100$ reflects the
    competition between the attractive and repulsive contributions.}
    \label{fig:cross_section_flow}
\end{figure}

\Cref{fig:cross_section_flow} shows the ratio 
\begin{equation}
    \frac{\sigma_V[V_{\rm tot7}]}{\sigma_V[V_{\rm rep}]}\equiv\dfrac{\sigma_{\rm Total}}{\sigma_{\rm Pauli}}
\end{equation}
as a function of the hierarchy parameter
$\xi=m_Q/\Lambda_d$ at $v_{\rm dwarf}=30~{\rm km/s}$. For
$N\lesssim75$, the full interaction produces pronounced resonant-like enhancements relative to the repulsive-only result.  The position of these enhancements
shifts towards larger $\xi$ as $N$ increases, and their shape becomes
broader, but the qualitative enhancement remains present over this entire range. For $N\sim100$, a different regime becomes apparent: the attractive and repulsive contributions  become comparable, resulting in a broad enhancement of the cross section over the
repulsive-only result. Increasing $N$ further strengthens the relative
importance of the repulsive core, and the enhancement is progressively
reduced, as illustrated by the $N=125$ and $150$ curves. In particular, at \(N=150\) we also observe an anti-resonance around \(\xi\sim10\), where the attractive and repulsive contributions nearly cancel in the scattering phase shift, resulting in a strongly suppressed cross section.

\begin{figure}[t]
    \centering
    \includegraphics[width=1\linewidth]{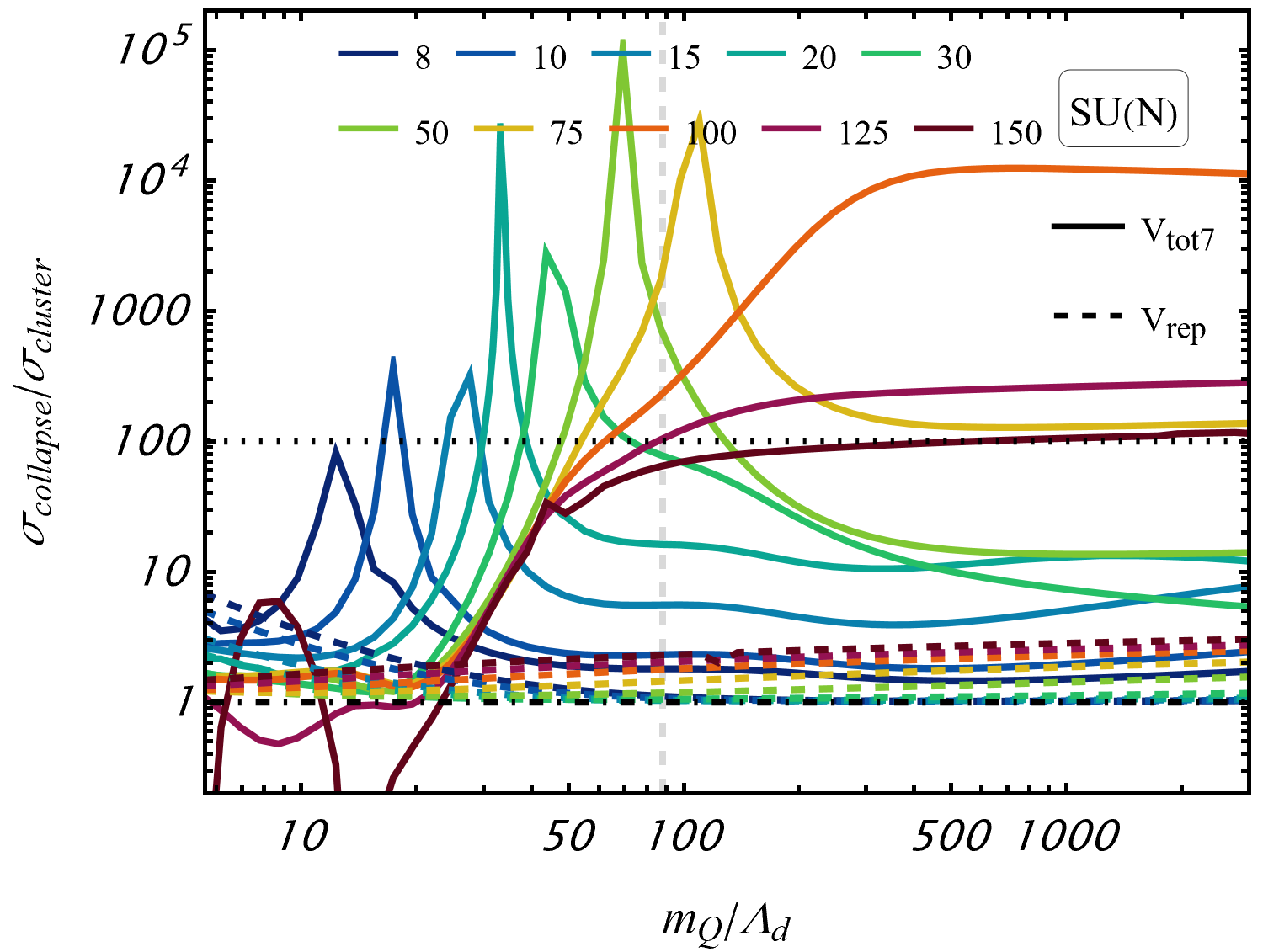}
    \caption{Ratio of the viscosity self-interaction cross section at
    $v_{\rm collapse}=5~{\rm km/s}$ to that at
    $v_{\rm cluster}=1500~{\rm km/s}$ as a function of
    $\xi=m_Q/\Lambda_d$ for increasing values of $N$ from 8 to 150.
    Solid curves are obtained using the full potential $V_{\rm tot}$ with $x_{\rm in}=7$,
    while dashed curves use the repulsive potential $V_{\rm rep}$ alone. The horizontal dot-dashed line
    indicates a ratio of unity, while the horizontal dotted line indicates
    a ratio of $100$, representative of the large low-velocity enhancement
    relevant for the gravothermal phenomenology considered in the main
    text. }
    \label{fig:cross_section_velocity_ratio7_core_colapse}
\end{figure}

The effect of this competition on the velocity dependence is shown in
\cref{fig:cross_section_velocity_ratio7_core_colapse}, where we consider the ratio\footnote{The condition that the baryon remains unresolved is gradually violated as $N$ increases, since $a_B k_{\rm max}\sim v_{\rm max}N/\tilde\alpha_d$. Although this does not constitute a strict validity condition for our computations, it indicates that the precise small-radius physics becomes increasingly important and that uncertainties associated with our short-distance regularization grow with $N$.}

\begin{equation}
    \frac{\sigma_V(v_{\rm collapse})}
         {\sigma_V(v_{\rm cluster})} \equiv\dfrac{\sigma_{\rm collapse}}{\sigma_{\rm cluster}},
\end{equation}
with $v_{\rm collapse}=5~{\rm km/s}$ and
$v_{\rm cluster}=1500~{\rm km/s}$. The solid curves use the full potential
$V_{\rm tot}$, while the dashed curves are obtained using the repulsive
potential $V_{\rm rep}$ alone. The latter remain close to a velocity-independent, geometric-like behavior, demonstrating that the large
velocity dependence of the full calculation does not originate from the
repulsive core itself. Instead, it results from the competition between the
attractive interaction and the repulsive core.

For $N\lesssim75$, the solid curves retain pronounced enhancements over the
cluster-scale cross section, associated with the resonant-like structures
seen in \cref{fig:cross_section_flow}. Around \(N\sim100\), the increasing repulsive contribution becomes sufficiently large to compete with the attraction over an extended range of \(\xi\), producing a broad enhancement.
This behavior arises because the perturbative attractive contribution is independent of \(\xi\) and the nonperturbative contribution becomes negligible for large $\xi$, so that increasing \(N\) can bring the repulsion into close competition with the attractive contribution over an extended range of \(\xi\). At still larger $N$,
the repulsive contribution becomes increasingly dominant and the low-velocity enhancement is reduced, as illustrated by the $N=125$ and $150$ curves.

Taken together, these results show that the velocity dependence of the
heavy-baryon self-interaction is controlled by the competition between the
attractive and repulsive components of the potential. For
$N\lesssim75$, this competition produces pronounced resonant-like
enhancements whose location moves to larger $\xi$ as $N$ increases. Around
$N\sim100$,the repulsion becomes sufficiently strong to compete with the attraction, producing a broad enhancement. For larger \(N\), the increasing strength of the repulsive interaction progressively drives the system towards a more geometric-like regime, reducing the velocity dependence. Thus, while the precise location and shape of the enhancement
depend on $N$, the qualitative mechanism persists over a substantial range
of moderately large $N$. At \(\xi\gg100\), the ratio of cross sections at collapse and cluster velocities instead approaches a plateau beyond the enhancement region. This asymptotic value provides a direct probe of \(N\), provided that a large \(\xi\gg100\) is independently established; the current observational targets considered here favor \(75\lesssim N\lesssim150\).

We restrict our numerical study to $N\lesssim150$. At larger \(N\), the rapidly increasing repulsive potential at distances of order the baryon size, combined with the large dimensionless kinetic term required at cluster velocities, makes the radial Schrödinger equation increasingly difficult to solve numerically. We therefore do not attempt to draw quantitative conclusions
about the detailed behavior at $N\gtrsim150$. The results above are sufficient to
establish the qualitative evolution from resonant-like behavior at
moderately large $N$, through the broad enhancement around $N\sim100$, to
the increasingly repulsion-dominated regime at the largest values studied.

\end{document}